\documentclass[12pt]{article}
\usepackage{psfig,epsf}
\usepackage{a4wide,graphicx}
\usepackage{cite}

\def\MeV{\textrm{ MeV}}
\def\GeV{\textrm{ GeV}}
\def\J{J/\Psi}
\newcommand{\be}{\begin{equation}}
\newcommand{\ee}{\end{equation}}
\newcommand{\ba}{\begin{eqnarray}}
\newcommand{\ea}{\end{eqnarray}}
\begin{document}

\title{Unitary chiral dynamics in $J/\Psi\to VPP$ decays and the role of
scalar mesons}

\author{
L.~Roca$^1$, J.~E.~Palomar$^1$, E.~Oset$^1$ and H.~C.~Chiang$^2$\\
{\it {\small $^1$Departamento de F\'{\i}sica Te\'orica and IFIC,
Centro Mixto Universidad de Valencia-CSIC,}} \\ 
{\it {\small Institutos de
Investigaci\'on de Paterna, Aptdo. 22085, 46071 Valencia, Spain}}\\ 
{\it {\small $^2$Institute of High Energy Physics, Chinese Academy of Sciences, Beijing
100039, China}}
}

\date{\today}

\maketitle 

\begin{abstract} 

We make a theoretical study of  the $\J$ decays into
$\omega\pi\pi$, $\phi\pi\pi$, $\omega K \bar{K}$ and $\phi
K\bar{K}$ using the techniques of the chiral unitary approach
stressing the important role of the scalar resonances
dynamically generated through the final state interaction of the
two pseudoscalar mesons. We also discuss the importance of new
mechanisms with intermediate exchange of vector and axial-vector
mesons and the role played by the OZI rule in the $\J\phi\pi\pi$
 vertex, quantifying its effects. The results nicely
reproduce the experimental data for the invariant mass
distributions in all the channels considered.

\end{abstract}

\section{Introduction}

The $\J$ decay into a pseudoscalar meson pair and a vector meson
has been claimed to be one of the most suited reactions to study
the long controversial nature of the scalar mesons, and much
work in this direction has been done both theoretically
\cite{Ishida,Pennington,Uehara:2002wh,OllerJ,Janssen:1994wn,Escribano:2002iv}
and experimentally \cite{WuBES,Augustin,Falvard,Lockman}. 
The nature of the scalar mesons is controversial and the
interpretation as $q\bar{q}$ mesons or as meson-meson molecules
has mainly centered the discussion  \cite{kyoto}.  When trying
to extract the physical properties of the scalar resonances from
experimental data, one has to be extremely careful in fitting
the theoretical models to the data since the 'bumps' or 'peaks'
in the invariant mass distributions are much influenced by the
particular dynamics of the production mechanisms. For instance,
the $f_0(980)$ peak in the $\phi\to\pi^0\pi^0\gamma$ decay is
distorted with respect to its shape in other reactions, because
gauge invariance of the production mechanisms introduces a
factor of the photon momentum which vanishes at the highest
$\pi\pi$ invariant mass, and grows fast as this mass decreases
passing through the $f_0(980)$ peak
\cite{Achasov:1987ts,Boglione:2003xh,phidecay}. Other sources of
distortion appear due to  interferences with other mechanisms or
non trivial effects due to the proximity of thresholds
\cite{phidecay,Li:2003zi}.

 Using the $\J\to \phi(MM)$ decay, the
authors of \cite{Pennington} found the $f_0(980)$ meson to have
a pole structure different to a $K\bar K$ molecule in contrast
to the findings of \cite{Janssen:1994wn}. In
\cite{Escribano:2002iv} the authors found that only a pole in
the second Riemann sheet was necessary to describe the
$\J\to\phi(\pi\pi,K\bar K)$ data. Concerning the $\sigma$ meson,
the $\pi\pi$ mass distribution of the  experimental $\J
\to\omega\pi\pi$ decay clearly shows an enhancement at around
$500\MeV$, which has been tried to be explained as a genuine
$\sigma$ meson described with Breit-Wigner shapes 
\cite{Augustin,Ishida,WuBES}.

 In the last years, a chiral unitary coupled channel approach  
\cite{Oller:1998hw,Oller:1998zr,Oller:2000ma} has proved to be
successful in describing meson-meson interactions in all
channels up to energies $\sim1.2\textrm{ GeV}$, far beyond the
natural limit  of applicability of the standard Chiral
Perturbation Theory (ChPT), which is $\sim500\MeV$ where the
pole of the lightest resonance, the $\sigma$ meson, appears. In
\cite{Oller:1998hw} the inverse amplitude method in coupled
channels is used while in \cite{Oller:1998zr} the $N/D$ unitary
method is exploited and it is shown to be equivalent to a
resummation of the loops implemented in a  coupled channel
Bethe-Salpeter equation using as kernel the lowest order ChPT
Lagrangian \cite{npa}. In this approach the scalar mesons, which
are a matter of concern in the present work, rise up naturally
as dynamically generated resonances, in the sense that, without
being included as explicit degrees of freedom, they
appear as poles in the s-wave meson-meson scattering amplitudes.
Concerning the $\J$ decays of interest in the present work, the
chiral unitary approach was used in Ref.~\cite{OllerJ} for these
$\J$ decays in order to evaluate the scalar form factor. In
Ref.~\cite{Uehara:2002wh} a similar technique was used to
implement the meson-meson rescattering in the same processes.

On the other hand, as pointed out in Ref.~\cite{OllerJ}, the
data on these $\J$ decays can be used to quantify the violation
of the second order Okubo-Zweig-Iizuka (OZI) rule in the
$0^{++}$ sector. In Ref.~\cite{OllerJ}, it was found that a
sizeable violation of the OZI rule was necessary to describe the
data, in agreement with the arguments of
Refs.~\cite{Stern,Isgur:2000ts} where the OZI rule violation in
the scalar sector is justified.

Given the controversy in the explanation of the  nature of the
scalar mesons from these $\J$ decays and the extraction of their
'physical' properties from the experimental data, the aim of the
present work is to make a consistent and comprehensive
description of the $\J\to VPP$ decays, including all the
mechanisms able to influence the region of pseudoscalar pair
invariant masses up to $\sim 1.1\textrm{ GeV}$, addressing the
main problems described above concerning the role played by the
scalar mesons and the OZI rule. First of all, in
Section~\ref{sec:chiral_loops}, we will address the same 
mechanisms used in  Ref.~\cite{OllerJ}, essentially the effect
of meson rescattering using techniques of the chiral unitary
approach, on top of the tree level $\J\to VPP$ amplitude
provided by a phenomenological local Lagrangian. We use $SU(3)$
arguments to relate the different channels and show the
equivalence to the formalism of Ref.~\cite{OllerJ}
and its relation to the OZI rule violation. In
Sections~\ref{sec:VMDtree} and  \ref{sec:VAloops} we explain
other mechanisms with sequential exchange of vector and
axial-vector mesons and, as an important novelty, the
meson-meson final state interaction. For the vertices needed in
these mechanisms we use previous Lagrangians
\cite{bramon5,escribano,axials} and propose new ones for those
involving the $\J$ meson. In the Results section we make a
thorough study of the role played by the scalar mesons and the
OZI rule in these decays and compare our results with
experimental data for $\omega\pi\pi$, $\phi\pi\pi$,
$\omega K\bar K$ and $\phi K\bar K$  from DM2
\cite{Augustin,Falvard}, MARK-III \cite{Lockman} and the recent
BES \cite{WuBES} experiments.

\section{The model for $J/\Psi\to VPP$ decay}

We proceed to construct our model by addressing different
mechanisms which, by analogy to other physical processes
previously studied, can significantly contribute to the
$J/\Psi\to VPP$ decays up to $PP$ invariant mass of around
$1.2\textrm{ GeV}$, where the scalar resonances $\sigma$ and
$f_0(980)$ play a very important role. Following the framework
of the chiral unitary approach, we will implement the final
meson-meson state interaction in order to generate dynamically
the scalar resonances involved. 

First of all let us define the nomenclature we will use
 for the kinematics along this work:
the decay we are considering is 
\begin{equation}
J/\Psi(\epsilon^*,q^*)\to P_1(p_1) + P_2(p_2) + V(\epsilon,q)
\end{equation}
with $\epsilon^*$ and $\epsilon$ the polarization vectors of the $\J$ and
the final vector meson respectively. The expression of the differential
decay width with respect to the invariant mass of the two pseudoscalars 
in the $\J$ rest frame can be evaluated as
\begin{equation}
\frac{d\Gamma}{dM_I}=\frac{M_I}{64\pi^3M^2}
\int_{m_1}^{M-\omega_q-m_2}d\omega_1
\overline{\sum}|t|^2\Theta(1-\cos\bar{\theta}^2)
\end{equation}
where $M$ is the $\J$ mass, $M_I$ is the invariant mass of the two
pseudoscalars, $\omega_i$ the on-shell energy
of the corresponding particle, $\Theta$  the step function and  
$\cos\bar{\theta}=\frac{(M-\omega_q-\omega_1)^2-m_2^2-
|\vec{q}|^2-|\vec{p_1}|^2}{2|\vec{q}| |\vec{p_1}|}$, where
$\bar{\theta}$ is the angle between $\vec{p}_1$ and $\vec{q}$.
The $t$-matrix can be expressed as
\be
t\equiv\epsilon^*_\mu\epsilon_\nu t^{\mu\nu}
\ee
and therefore the polarization sum is
\be
\sum|t|^2=\sum_{\mu\mu'\nu\nu'}
\left(-g_{\mu{\mu}'}+\frac{q^*_\mu q^*_{\mu'}}{M^2}\right)
\left(-g_{\nu{\nu}'}+\frac{q_\nu q_{\nu'}}{M_V^2}\right)
t^{\mu\nu}{t^*}^{\mu'\nu'}
\ee

In the next subsections we evaluate the different contributions
to the amplitude.

\subsection{The $J/\Psi PPV$ vertex with meson loops
\label{sec:chiral_loops}}

\begin{figure}
\centerline{\protect\hbox{
\psfig{file=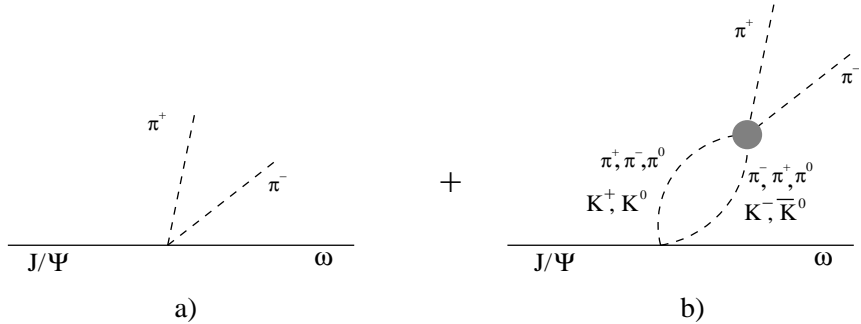,width=0.7\textwidth,silent=}}}
\caption{Diagrams with direct $\J VPP$ vertex at tree level, a), and with
the iterated meson loops, b). }
\label{fig:chiral_loops}
\end{figure}

The first mechanisms to be considered are those involving a
direct coupling of the $\J$ to the two pseudoscalars and the
vector, implementing the final state interaction of the
pseudoscalars pair, as  is depicted in
Fig.~\ref{fig:chiral_loops} for the $\J\to\omega\pi^+\pi^-$
channel.

The thick dot in Fig.~\ref{fig:chiral_loops} means that one is
considering the full $\pi\pi(K\bar{K})\to \pi^+\pi^-$
$t$-matrix, involving the loop resummation of the Bethe-Salpeter
equation of Ref.~\cite{npa} and no just the lowest order
$\pi\pi(K\bar{K})\to \pi^+\pi^-$ amplitude. Actually this loop
resummation is what dynamically generates the scalar resonances,
in the sense that the scalars are not explicitly included in
the  model but they appear naturally as poles in the meson-meson
scattering amplitude \cite{npa}.

For the evaluation of these diagrams we need to know the vertex
accounting for the transition of the $\J$ into a vector meson
and a pseudoscalar meson pair. We will consider the pseudoscalar
pair having the vacuum quantum numbers, $J^{PC}=0^{++}$, since,
as mentioned in the introduction, the $\J$ decays considered in
the present work are dominated by the scalar sector at the
energies that we are interested in. Similarly to what is done in
Ref.~\cite{OllerJ} we write a contact term of the form
\be
\Psi_\mu V^{\mu} \phi\phi';
\label{eq:contact}
\ee
where $\Psi_\mu$, $V^{\mu}$ and $\phi\phi'$ are the fields of 
$\J$, vector and pseudoscalar mesons respectively,
which according to
Ref.~\cite{OllerJ} is accurate enough since any other possible
structures containing derivatives of the fields provide small
momentum dependences, given the large $\J$ mass. 

The relative weights between the different channels can be
obtained using $SU(3)$ arguments in the following way: the
physical meson-meson states can be decomposed in terms of the
basis states of the singlet ($\bar S_1$), symmetric-octet
($\bar S_8^s$) and antisymmetric-octet ($\bar S_8^a$)
 representations of
$SU(3)$ by means of the $SU(3)$ Clebsch-Gordan coefficients.
This decomposition gives:

\ba
\nonumber
|K\bar{K}>&=&-\frac{1}{2\sqrt{2}}
|\bar S_1>-\frac{1}{2\sqrt{5}}|\bar S_8^s>-\frac{1}{2}
|\bar S_8^a>          \\
|\pi\pi>&=&-\frac{1}{2\sqrt{2}}|\bar S_1>
+\frac{1}{\sqrt{5}}|\bar S_8^s>,
\label{eq:SU3CG}
\ea
where we have also used a minus sign in the phase of the $\pi^+$
and $K^-$ states.

From Eq.~(\ref{eq:SU3CG}), and with the Wigner-Eckart theorem in
mind and taking into account that the $\J$ can be considered as
a $SU(3)$ singlet, it is direct to write the following matrix
elements:

\ba
\nonumber
<V_1|t|K\bar{K}>&=&
-\frac{1}{2\sqrt{2}}<V_1|t|\bar S_1>
\equiv -\frac{1}{2\sqrt{2}}t_1\\ \nonumber
<V_8^s|t|K\bar{K}>&=&-\frac{1}{2\sqrt{5}}<V_8^s|t|\bar S_8^s>
\equiv -\frac{1}{2\sqrt{5}}t_8^s\\ \nonumber
<V_1|t|\pi\pi>&=&-\frac{1}{2\sqrt{2}}<V_1|t|\bar S_1>
\equiv -\frac{1}{2\sqrt{2}}t_1\\
<V_8^s|t|\pi\pi>&=&-\frac{1}{\sqrt{5}}<V_8^s|t|\bar S_8^s>
\equiv \frac{1}{\sqrt{5}}t_8^s,
\label{eq:redt}
\ea

\noindent
where $t_1$ and $t_8^s$ are the reduced matrix elements which we
will consider as unknown coefficients. Actually, $t_1$ and
$t_8^s$ will be the only two free parameters in
all the model described along the present work.   Given the
symmetry $K\bar K \leftrightarrow \bar K K$ in s-wave and that
$\bar K K$ has a coefficient $+1/2$ for the $|\bar S_8^a>$
state, instead of $-1/2$ in Eq.~(\ref{eq:SU3CG}) for $K\bar K$,
the matrix elements with the antisymmetric octet state vanish.

On the other hand, considering ideal mixing between the $V_8$
and $V_1$ states (we omit the index "$s$" for symmetric in what
follows), we can write the following decomposition in terms of
the physical $\phi$ and $\omega$ meson states:

\be
V_1=\sqrt{\frac{2}{3}}\omega+\frac{1}{\sqrt{3}}\phi,
\qquad   V_8=\frac{1}{\sqrt{3}}\omega-\sqrt{\frac{2}{3}}\phi.
\label{eq:wphimix}
\ee

Using Eqs.~(\ref{eq:redt}) and (\ref{eq:wphimix}) and including also the 
polarization vectors of the $\J$, $\epsilon^*$, and the vector meson,
$\epsilon$,
 the amplitudes for the
different contact terms involving the $\J$, one vector and two
pseudoscalars needed to evaluate the diagrams of
Fig.~\ref{fig:chiral_loops}, are:

\ba
\nonumber
t_{\J \phi
K\bar{K}}&=&-\frac{1}{\sqrt{6}}
\left(\frac{1}{2}t_1-\frac{1}{\sqrt{5}}t_8\right)\,\epsilon^*\cdot\epsilon \\ \nonumber
t_{\J \omega
K\bar{K}}&=&-\frac{1}{2\sqrt{3}}
\left(t_1+\frac{1}{\sqrt{5}}t_8\right)\,\epsilon^*\cdot\epsilon \\ \nonumber
t_{\J \phi
\pi\pi}&=&-\frac{1}{\sqrt{6}}
\left(\frac{1}{2}t_1+\frac{2}{\sqrt{5}}t_8\right)\,\epsilon^*\cdot\epsilon \\
t_{\J \omega
\pi\pi}&=&-\frac{1}{2\sqrt{3}}
\left(t_1-\frac{2}{\sqrt{5}}t_8\right)\,\epsilon^*\cdot\epsilon.
\label{eq:tvertex}
\ea

At this point it is worth mentioning the implications of the OZI
rule in this decays. A thorough study and explanation of the
role played by the OZI rule in the $\J$ decays into a $\phi$
meson and two pseudoscalars was done in Ref.~\cite{OllerJ}. For
the purpose of the present work it is enough to point out that,
due to the non-existence of direct quark line connexion between
the strange and up or down quarks, the $\J\to\phi\pi\pi$ decay
is suppressed to next order in $\alpha_s$ with respect to the
other channels.  Should the OZI rule be  exact,
 $t_{\J\phi\pi\pi}$
  would be zero, implying that $t_1=(-4/\sqrt{5})t_8$.
Sizeable deviations from this numerical relation would point out
to a necessary deviation from the OZI rule. This is something to
be expected since this rule is only well founded for the large
$N_c$ limit of QCD and the $0^{++}$ sector is not well described
in this limit. In Ref.~\cite{OllerJ} a different approach was
followed parametrizing the Lagrangian in terms of a scalar
source $S$ which plays the role of the pseudoscalar pairs with
scalar quantum numbers in our model, and this scalar is written
in terms of quark fields as
\be
\bar\Psi_\mu V^\mu S
\ee
and
 \be
  S\equiv\bar s s +
\lambda_{\phi}\frac{1}{\sqrt{2}}(\bar u u+\bar d d).
\label{eq:deflambda}
 \ee 
In this way, the $\lambda_\phi$
parameter accounts for the relative weight of the non-strange
quark content of the scalar sources. Therefore it quantifies the
OZI rule violation in the case of two pions in the scalar
channel connected to the $\phi$ and the $\J$. In
Ref.~\cite{OllerJ} the following relations of
the quark-antiquark operators in terms  of the meson-meson
fields were obtained from the mass term of the lowest order
ChPT Lagrangian, (see that reference for details):

\ba  
\bar{u}u&=&-f^2 B_0  
\left[1-\frac{1}{f^2}\left(\pi^+ \pi^- + K^+ K^- + \frac{(\pi^0)^2}{2} + \frac{\eta_8^2}{6} +   
\frac{\pi^0 \eta_8}{\sqrt{3}}\right)+...\right] \nonumber \\  
\bar{d}d&=&-f^2 B_0  
\left[1-\frac{1}{f^2}\left(\pi^+ \pi^- + K^0 \overline{K}^0 + \frac{(\pi^0)^2}{2} +   
\frac{\eta_8^2}{6} - \frac{\pi^0 \eta_8}{\sqrt{3}}\right)+...\right] \nonumber \\  
\bar{s}s&=&-f^2  
B_0\left[1-\frac{1}{f^2}\left(K^+K^- + K^0 \overline{K}^0 +\frac{2}{3}\eta_8^2\right)+...  
 \right]
 \label{eq:qqMM}    
\ea  

\noindent  
where the dotted points represent higher order in the meson fields. The
small non-strange content of the pions is clearly manifest in the last of
these equations since the pion fields would appear at higher orders in
the meson fields. Introducing the octet and singlet scalar sources, $V_8$
and $V_1$ respectively, the Lagrangian for the contact term interaction
can be expressed as

\be
\hat{g}\,\Psi_\mu(V_8^\mu S_8+\nu V_1^\mu  S_1)
\label{eq:Loller}
\ee
with $\nu$ an unknown parameter accounting for the relative weight between
the singlet and octet couplings, which in Ref.~\cite{OllerJ}
is related to the
$\lambda_\phi$ parameter of Eq.~(\ref{eq:deflambda}).

\noindent
One can introduce the scalar sources $S_\omega$ and $S_\phi$ in an 
analogous way
to Eq.~(\ref{eq:wphimix}):
\be
S_1=\sqrt{\frac{2}{3}} S_\omega +\frac{1}{\sqrt{3}}S_\phi,
\qquad   S_8=\frac{1}{\sqrt{3}}S_\omega-\sqrt{\frac{2}{3}}S_\phi,
\label{eq:wphimix2}
\ee
where in a quark model language, consistently with the transformation properties under
SU(3), we can write:
\be
\label{eq:ssq}
S_\phi=\bar{s}s\;\;\;\;\hbox{and}\;\;\;\; S_\omega=\frac{1}{\sqrt{2}}(\bar{u}u+\bar{d}d)
~.
\ee

Combining Eqs.~(\ref{eq:qqMM}), (\ref{eq:Loller}), (\ref{eq:wphimix2})  and 
(\ref{eq:ssq}) one obtains the following amplitudes for the
different contact terms involving the $\J$, one vector and two
pseudoscalars:

\ba
\nonumber
t_{\J \phi
K\bar{K}}&=&-\frac{\tilde{g}}{3}(2\nu+1)\,\epsilon^*\cdot\epsilon \\ \nonumber
t_{\J \omega
K\bar{K}}&=&-\frac{\tilde{g}}{3\sqrt{2}}(4\nu-1)\,\epsilon^*\cdot\epsilon \\ \nonumber
t_{\J \phi
\pi\pi}&=&-\frac{2\tilde{g}}{3}(\nu-1)\,\epsilon^*\cdot\epsilon \\
t_{\J \omega
\pi\pi}&=&-\frac{\sqrt{2}\tilde{g}}{3}(2\nu+1)\,\epsilon^*\cdot\epsilon.
\label{eq:tvertexOll}
\ea
with $\nu=\frac{\sqrt{2}+2\lambda_\phi}{\sqrt{2}-\lambda_\phi}$ and
$\widetilde{g}\equiv \hat{g}B_0$ where $B_0$ is the 
constant appearing in the mass term of the chiral Lagrangian
\cite{Gasser:1984gg}. In these amplitudes 
$\widetilde{g}$ and $\nu$ (or equivalently the OZI violation parameter
$\lambda_\phi$) are the two free parameters. Note that the exact
accomplishment of the OZI rule would require $\lambda_\phi=0$ (see
Eq.~(\ref{eq:deflambda})),
and consequently $\nu=1$, and therefore it would imply the third
equation of Eqs.~(\ref{eq:tvertexOll}) to be zero.

Comparing Eqs.~(\ref{eq:tvertex}) and (\ref{eq:tvertexOll}) it is
immediate to see the equivalence between the two different
treatments of the $SU(3)$ symmetry by writing

\be
t_1=4\sqrt{\frac{2}{3}}\widetilde{g}\nu\qquad,\qquad
t_8=-\sqrt{\frac{10}{3}}\widetilde{g}.
\label{eq:t1t8}
\ee

The treatment of Eqs.~(\ref{eq:contact}) to (\ref{eq:tvertex})
is very intuitive and easy while the one from
Eqs.~(\ref{eq:deflambda}) to (\ref{eq:t1t8}) has the virtue of
expressing the amplitudes directly in terms of the OZI rule
violating parameter. In order to favour comparison  of our
results with those of Ref.~\cite{OllerJ}, we shall adopt their
nomenclature, using  $\widetilde{g}$ and $\lambda_\phi$ as free
parameters  instead of $t_1$ and $t_8$.

 All this said, we can already write the amplitudes for the
diagrams depicted in Fig.~\ref{fig:chiral_loops} for all the
channels of concern in the present work:

\ba
\nonumber
t_{\J\to\omega\pi^+\pi^-}=-\widetilde{g}\epsilon^*·\epsilon
\frac{\sqrt{2}}{3} \left[\frac{4\nu-1}{\sqrt{3}}G_{KK}t_{KK,\pi\pi}^{I=0}
+(1+2\nu)
\left(1+G_{\pi\pi}t_{\pi\pi,\pi\pi}^{I=0}\right)\right] \\
\nonumber
t_{\J\to\phi\pi^+\pi^-}=-\widetilde{g}\epsilon^*·\epsilon
\frac{2}{3}\left[\frac{1+2\nu}{\sqrt{3}}G_{KK}t_{KK,\pi\pi}^{I=0}
+(\nu-1)
\left(1+G_{\pi\pi}t_{\pi\pi,\pi\pi}^{I=0}\right)\right] \\ \nonumber
t_{\J\to\omega K^+K^-}=-\widetilde{g}\epsilon^*·\epsilon
\frac{1}{3\sqrt{2}}\left[(4\nu-1)(1+G_{KK}t_{KK,KK}^{I=0})
+ \sqrt{3}(1+2\nu)
G_{\pi\pi}t_{\pi\pi,KK}^{I=0}\right] \\
t_{\J\to\phi K^+K^-}=-\widetilde{g}\epsilon^*·\epsilon \frac{1}{3}
\left[(1+2\nu)(1+G_{KK}t_{KK,KK}^{I=0})
+ \sqrt{3}(\nu-1)
G_{\pi\pi}t_{\pi\pi,KK}^{I=0}\right]
\label{eq:tchiral_loops}
\ea

In Eq.~(\ref{eq:tchiral_loops}) $G_{\pi\pi}$ and $G_{KK}$ are
the ordinary two meson loop functions regularized by means of a
cutoff of the order of $1\GeV$, $t^{I=0}_{MM,M'M'}$ are the
isospin zero  $MM\to M'M'$ transition amplitudes, accounting for
the resummation of the iterated loops, evaluated with the
techniques of the chiral unitary approach of Ref.~\cite{npa}.
Note that the meson-meson scattering amplitudes have been
factorized on shell out of the loops as was justified in
Ref.~\cite{npa}. In Eq.~(\ref{eq:tchiral_loops}) we have also
taken into account that 

\ba
\nonumber
<\pi^+\pi^- + \pi^-\pi^+ + \pi^0\pi^0 |t_m|\pi^+\pi^->
&=&2t_{\pi\pi,\pi\pi}^{I=0} \\ \nonumber
<K^+ K^- + K^0 \bar{K}^0 |t_m|\pi^+\pi^-> 
&=&\frac{2}{\sqrt{3}}t_{KK,\pi\pi}^{I=0} \\ \nonumber
<K^+ K^- + K^0 \bar{K}^0 |t_m|K^+ K^->&=&t_{KK,KK}^{I=0} \\
<\pi^+\pi^- + \pi^-\pi^+ + \pi^0\pi^0 |t_m|K^+K^->
&=&\sqrt{3}t_{KK,\pi\pi}^{I=0} 
\ea
where the unitary normalization for the states of 
Ref.~\cite{npa} has been used.

\subsection{Sequential vector and axial-vector meson exchange
 mechanisms: tree level
\label{sec:VMDtree}}

Previous works on $\rho$, $\omega$ \cite{escribano,palomar}
and $\phi$ \cite{phidecay,achasovlast,lucio} decays into  two
pseudoscalars and one photon, showed that the mechanisms
where the initial vector meson decays into a pseudoscalar and
another vector meson, and this latter one decays itself into a
pseudoscalar and one photon, play an important role. The strong
analogy with the decays studied in the present work suggests to
study the role played in these $\J$ decays by these kind of
mechanisms, not previously considered in
other works. On the other hand, the
strong meson-meson scattering of the final pseudoscalar mesons,
specially in the scalar channel,
which we will discuss in subsection
\ref{sec:VAloops}, makes definitively necessary their study.

In what follows we will explicitly discuss the
$\J\to \omega\pi^+\pi^-$
channel, referring to the Appendix for the
formulae of the other channels since they are analogous.

\begin{figure}[tbp]
\centerline{\hbox{\psfig{file=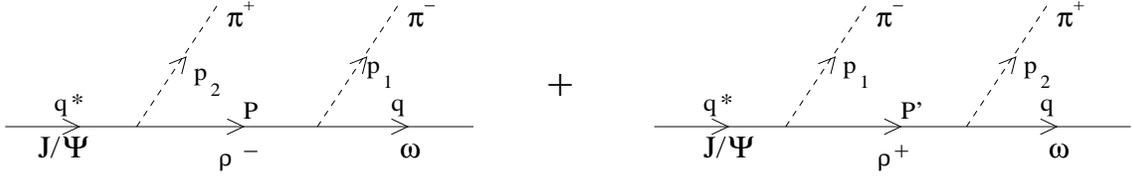,width=15cm}}}
\caption{\rm 
Diagrams for the tree level mechanisms with sequential vector meson
exchange.}
 \label{fig:VMD_tree_w}
\end{figure}

In Fig.~\ref{fig:VMD_tree_w} the two allowed diagrams at tree
level  for the sequential vector meson exchange are depicted,
including the notation for the momenta.

For the evaluation of the vector-vector-pseudoscalar (VVP) vertex we use
the same Lagrangians as in \cite{bramon5,escribano}

\be
  \label{eq:LVVP}
{\cal L}_{VVP} = \frac{G}{\sqrt{2}}\epsilon^{\mu \nu \alpha \beta}\langle
\partial_{\mu} V_{\nu} \partial_{\alpha} V_{\beta} P \rangle
\ee
where $\langle\rangle$ means $SU(3)$ trace,
 $G=0.016\textrm{ MeV}^{-1}$ and  $V$ $(P)$ are the usual
 vector (pseudoscalar) $SU(3)$ matrices.
 In an analogous way, we can use the following
  Lagrangian for the  $\J VP$ vertex:
\be
 \label{eq:LJVP}
{\cal L}_{\Psi VP} = \frac{\overline{G}}{\sqrt{2}}
\epsilon^{\mu \nu \alpha \beta}\langle
\partial_{\mu} \Psi_{\nu} \partial_{\alpha} V_{\beta} P \rangle
=\frac{\overline{G}}{\sqrt{2}}
\epsilon^{\mu \nu \alpha \beta}
\partial_{\mu} \Psi_{\nu} \langle
\partial_{\alpha} V_{\beta} P \rangle.
\ee  
where in the last step the $\J$ field, $SU(3)$ singlet, is
factorized out of the $SU(3)$ trace.
From the experimental $\J\to VP$ decay widths from
the PDG \cite{pdg}, we obtain the numerical value of the
coupling constant with its error:
 $\overline{G}=(1.44\pm 0.05)\times 10^{-6}\MeV^{-1}$, with the overall
sign unknown. The uncertainties coming from the sign and the experimental
errors will be discussed in the Results section.

 With the Lagrangians of Eqs.~(\ref{eq:LVVP}) and (\ref{eq:LJVP}), the
 amplitude for the diagrams depicted in Fig.~\ref{fig:VMD_tree_w} reads
 
 \begin{equation}
t=-\frac{G\overline{G}}{\sqrt{2}}
\left[\frac{P^2\{a\}+\{b(P)\}}{M_{\rho}^2-P^2-iM_\rho
\Gamma_\rho(P^2)}
+\frac{P'^2\{a\}+\{b(P')\}}{M_{\rho}^2-P'^2-iM_\rho
\Gamma_\rho(P'^2)}
\right]
\label{eq:tVtree}
\end{equation}

\noindent
 where  $P=p_1+q$, $P'=p_2+q$ and
\begin{eqnarray} \nonumber
\{a\}&=&\epsilon^*\cdot\epsilon \ q^*\cdot q
- \epsilon^*\cdot q \ \epsilon\cdot q^* \\
\{b(P)\}&=&-\epsilon^*\cdot\epsilon \ q^*\cdot P \ q\cdot P
- \epsilon\cdot P \ \epsilon^*\cdot P \ q^*\cdot q
+ \epsilon^*\cdot q \ \epsilon\cdot P \ q^*\cdot P
+  \epsilon\cdot q^* \ \epsilon^*\cdot P \ q\cdot P
\nonumber
\end{eqnarray}

In Ref.~\cite{phidecay} the important role of the analogous
mechanisms considering the exchange of an axial-vector meson
with $J^{PC}=1^{+-}$ or $1^{++}$ was shown. In 
Table~\ref{tab:axials} we show the particles of these octets.
\begin{table}[h]
\begin{center}
\begin{tabular}{|c||c||c|c|}\hline 
$J^{PC}$ &$I=1$  &$I=0$ & $I=1/2$ \\ \hline \hline  
 $1^{+-}$  & $b_1(1235)$  & $h_1(1170)$, $h_1(1380)$
    &$K_{1B}$ \\ \hline  
 $1^{++}$  & $a_1(1260)$  & $f_1(1285)$, $f_1(1420)$
    &$K_{1A}$ \\ \hline     
\end{tabular}
\end{center}
\caption{Octets of axial-vector mesons.}
\label{tab:axials}
\end{table}
 In
addition one has to consider the mixture of the
$K_{1B}$ and $K_{1A}$ states to give the physical $K_1(1270)$
and $K_1(1400)$  states:
\begin{eqnarray} \nonumber 
K_1(1270)&=&\cos(\alpha) K_{1B} -i \sin(\alpha) K_{1A} \\
K_1(1400)&=&\sin(\alpha) K_{1B}+i \cos(\alpha) K_{1A}
\label{eq:mixK}
\end{eqnarray}
with $\alpha\simeq 45$~degrees\footnote{It is worth mentioning that
 in \cite{suzuki,axials} two
more possible solutions for the mixing angle between the I=1/2 members of
the axial-vector octets were found around $30$ and
$60$~degrees . This uncertainty will be taken into account in the
evaluation of the theoretical error band of the final results.}.

In the $\J\to\omega\pi^+\pi^-$ decay, the mechanisms are
equivalent to those of Fig.~\ref{fig:VMD_tree_w} substituting
the intermediate $\rho$ meson by the $b_1(1235)$. The non
negligible contribution of these new mechanisms was already
pointed out  in Ref.~\cite{Uehara:2002wh}. For the evaluation of
these diagrams we need the couplings of the axial-vectors to one
vector and one pseudoscalar and the $\J$ to one axial vector and
one pseudoscalar.  For the first one we use the phenomenological
Lagrangian proposed in Ref.~\cite{axials} which successfully
describes the experimental branching ratios of one axial-vector
decay into one vector plus one pseudoscalar
and was used in \cite{phidecay} in radiative $\phi$ decay.
This Lagrangian is:

\be  \label{eq:LBLA}
{\cal L}_{A(B)VP}=D \langle V_{\mu\nu}\{B^{\mu\nu},P\}\rangle
-iF \langle V_{\mu\nu}[A^{\mu\nu},P]\rangle
\ee

\noindent
with $D=-1000\pm 120\MeV$; $F=1550\pm 150\MeV$ and
$B$ and $A$ are the $SU(3)$ matrices, in the tensor formalism of
\cite{ecker}, 
containing the octet of $1^{+-}$ and $1^{++}$ respectively. In this
tensor formalism  the fields 
$W_{\mu\nu}\equiv V_{\mu\nu}, \,B_{\mu\nu},\,A_{\mu\nu}$ are normalized
such that 
\begin{equation}
<0|W_{\mu\nu}|W;P,\epsilon>=
\frac{i}{M_W}\left[ P_\mu\,\epsilon_\nu(W)-P_\nu\, 
\epsilon_\mu(W)\right]
\end{equation}
In addition the propagators with the tensor fields are given
by \cite{ecker}
\begin{eqnarray} \label{eq:prop_4indices}
&<&0|T\{W_{\mu\nu}W_{\rho\sigma}\}|0>=i{\cal
D}_{\mu\nu\rho\sigma} =\\ \nonumber
&=&i\frac{M_W^{-2}}{M_W^2-P^2-i\epsilon}\left[g_{\mu\rho}\,
g_{\nu\sigma}\,(M_W^2-P^2)
+g_{\mu\rho}\,
P_\nu \, P_\sigma-g_{\mu\sigma}\,P_\nu
 \, P_\rho-(\mu \leftrightarrow \nu) \right].
\end{eqnarray}

For the vertex involving the $\J$, one axial and one
pseudoscalar  meson we can replace in Eq.~(\ref{eq:LBLA})
$V_{\mu\nu}$ by  the $\J$ field, $\Psi_{\mu\nu}$, where, since
$\Psi_{\mu\nu}$ can be considered as an $SU(3)$ singlet, it
factorizes out of the $SU(3)$ trace. Therefore we have

\ba  \nonumber
{\cal L}_{A(B)\Psi P}&=&\frac{\overline{D}}{2} \Psi_{\mu\nu}
\langle \{B^{\mu\nu},P\}\rangle
-i\overline{F} \Psi_{\mu\nu} \langle [A^{\mu\nu},P] \rangle \\
 &=&\overline{D} \Psi_{\mu\nu} \langle B^{\mu\nu}P \rangle
 \label{eq:LJLA}
 \ea

\noindent 
since  $\langle [A^{\mu\nu},P] \rangle =0$. In
Eq.~(\ref{eq:LJLA}) we have obtained that there is no direct
coupling of the $\J$ to the octet of $1^{++}$ axial-vector
mesons and one pseudoscalar, something that is in agreement with
the absence of experimental evidence of these decays \cite{pdg}.
This makes us confident in the phenomenological
reliability of the Lagrangian of Eq.~(\ref{eq:LJLA}). From the
PDG experimental values of the $\J$ decay into an axial-vector
meson and a pseudoscalar we obtain $\overline{D}=2.12\pm
0.24\MeV$, with an overall undetermined sign which will be
discussed in the Results section. 

With the Lagrangians of Eqs.~(\ref{eq:LBLA}) and
(\ref{eq:LJLA}),  the amplitude for the tree level sequential
mechanism with the exchange  of an axial-vector meson is

\begin{eqnarray}
\nonumber
t &=& \frac{4\sqrt{2}D\overline{D}}{M m_\omega M_b^2} 
\left[ (\epsilon^*\cdot\epsilon \ q^*\cdot q
- \epsilon^*\cdot q \ \epsilon\cdot q^*) \ + \ 
 \frac{1}{M_{b}^2-P^2-iM_b\Gamma_b(P^2)}  \cdot \right.\\ 
&\cdot& \left. \left(\epsilon^*\cdot\epsilon \ q^*\cdot P \ q\cdot P
+ \epsilon\cdot P \ \epsilon^*\cdot P \ q^*\cdot q
- \epsilon^*\cdot q \ \epsilon\cdot P \ q^*\cdot P
-  \epsilon\cdot q^* \ \epsilon^*\cdot P \ q\cdot P
\right)\right]
\label{eq:tAtree}
\end{eqnarray}

\noindent 
plus the crossed term, with $P'$ instead of $P$.

\subsection{Final 
state interaction in the sequential vector and axial-vector exchange
mechanisms \label{sec:VAloops}}

\begin{figure}[tbp]
\centerline{\hbox{\psfig{file=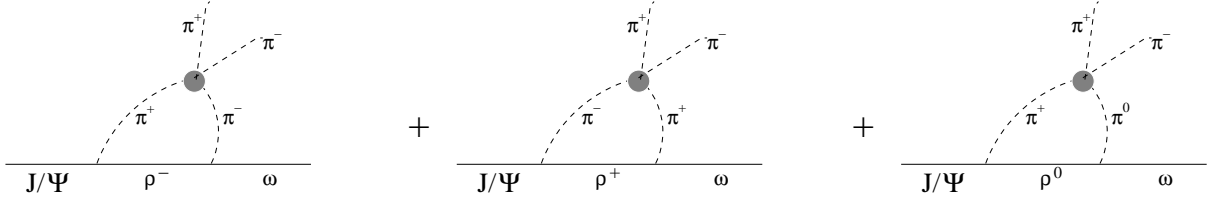,width=16cm}}}
\caption{\rm 
Sequential vector meson exchange diagrams with final state
interaction of pions}
\label{fig:VMD_loop_pions_w}
\end{figure}

\begin{figure}[tbp]
\centerline{\hbox{\psfig{file=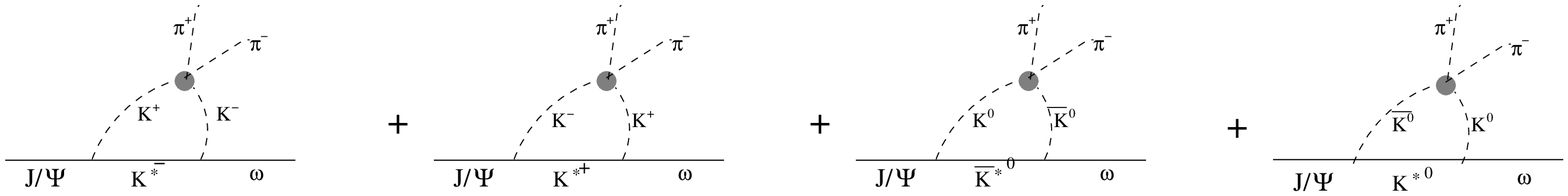,width=16cm}}}
\caption{\rm 
Sequential vector meson exchange diagrams with final state
interaction of kaons}
\label{fig:VMD_loop_kaons_w}
\end{figure}

Since the meson-meson interaction is strong in the region of
invariant masses relevant in the present reaction, specially in the scalar
channel, we next
consider the final state interaction of the two pions in the
sequential vector meson mechanism (see Fig.~\ref{fig:VMD_loop_pions_w})
but also with kaons in
the intermediate states (Fig.~\ref{fig:VMD_loop_kaons_w}). Again the thick dot
in Figs.~\ref{fig:VMD_loop_pions_w} and ~\ref{fig:VMD_loop_kaons_w}
means that one is considering the full meson-meson to $\pi\pi$
$t$-matrix, involving the loop resummation of the BS equation of
Ref.~\cite{npa} and not just the lowest order amplitude.

To evaluate these diagrams we need to calculate the loop functions
containing a vector and two pseudoscalar meson propagators. The
evaluation of these three meson loop functions was done in
Ref.~\cite{phidecay}. We summarize here the main steps:

given the
structure of the terms in Eq.~(\ref{eq:tVtree}) 
we must evaluate the loop integrals 

\begin{equation}   \label{eq:loop3mesons}
i\int\frac{d^4P}{(2\pi)^4} \,P^{\mu}P^{\nu}
\frac{1}{P^2-M_b^2+i\epsilon}
\,\frac{1}{(q^*-P)^2-m_1^2+i\epsilon}
\,\frac{1}{(q-P)^2-m_2^2+i\epsilon}
\end{equation}
where $m_1$ and $m_2$ are the masses of the pseudoscalar mesons of the
loops and $P$ is the momentum of the vector meson in the loop.

For simplicity we evaluate these integrals in the
reference frame where  the two meson system has zero momentum. In
this frame the $\J$ and $\omega$ trimomentum are the same,
$\vec{q}$. At the end we will boost back the 
$t^{\mu\nu}$-matrix to the
$\J$ rest frame where the phase-space integration and the polarization
vectors sum is done.

Given the momentum structure of Eqs.~(\ref{eq:tAtree})
and (\ref{eq:tVtree}), we need the following integrals, which for
dimensional reasons we write as

\begin{eqnarray}  \nonumber
i\int\frac{d^4P}{(2\pi)^4} P^0P^0 D_1 D_2 D_3 &=& I_0\\
\nonumber
i\int\frac{d^4P}{(2\pi)^4} P^0P^i D_1 D_2 D_3
 &=& \frac{q^i}{|\vec{q}|}I_1\\
i\int\frac{d^4P}{(2\pi)^4} P^iP^j D_1 D_2 D_3 
&=& \delta_{ij}I_a+\frac{q^iq^j}{\vec{q}\,^2}I_b
\label{eq:Is}
\end{eqnarray}
where $D_1$, $D_2$, $D_3$ are the three meson propagators of 
Eq.~(\ref{eq:loop3mesons}) and the momenta are in the dimeson rest frame.
 Analytical expressions for the $P^0$ integration in Eq.~(\ref{eq:Is})
 can be found in the
 Appendix of Ref.~\cite{phidecay}. The $d^3\vec{P}$ integral is evaluated
numerically by means of the same cut off, of the order of $1\,$GeV,
 which has been used
to regularize the two meson loop in the meson-meson
interaction.

With the structure of  Eq.~(\ref{eq:tVtree}) and the
substitutions of Eq.~(\ref{eq:Is}), one can already evaluate the
amplitude $t^{\mu\nu}$ for the diagrams of
Fig.~\ref{fig:VMD_loop_pions_w} which, after the proper boost
transformation to the $\J$ rest frame and after some
calculations, reads as

\begin{equation}
t^{\mu\nu} = \frac{G \overline{G}}{\sqrt{2}}
\,\bar{t}'^{\mu\nu}\,2 t_{\pi\pi,\pi\pi}^{I=0}
\end{equation}
with
\begin{equation}
\bar{t}'^{\mu \nu} = \left(\begin{array}{cccc} 
0 & 0 & 0 & 0 \\
\frac{2I_a}{M^2} |\vec q|({{q^*}^0}^2-{\vec{q}\,^2})({{q^*}^0}- {q^0})
 & \frac{2 I_a}{M^2} ({{q^*}^0}^2-{\vec{q}\,^2})
 ({{q^*}^0} {q^0}-{\vec{q}\,^2}) & 0 & 0 \\
0 & 0 & \bar{t}'_{22} & 0 \\
0 & 0 & 0 & \bar{t}'_{33} \\
\end{array}
\right)
\label{eq:tbar}
\end{equation}
where
$\bar{t}'_{22}=\bar{t}'_{33}\equiv {\vec{q}\,^2} I_0-|\vec q|(q^0+{q^*}^0)I_1-
({\vec{q}\,^2}-2q^0{q^*}^0 )I_a+q^0{q^*}^0I_b$.
The matrix $\bar{t}'^{\mu \nu}$ is the proper Lorentz tensor in
the $\J$ rest frame, although for convenience (since the functions
$I_i$ are evaluated in the dimeson rest frame) we write the 
$\bar{t}'^{\mu \nu}$ components in terms of the dimeson rest
frame momenta given by
\be
q^0=-\frac{M_I^2-M^2+m_\omega^2}{2M_I}\, ,
\quad q^{*0}=M_I+q^0=\frac{M_I^2+M^2-m_\omega^2}{2M_I}
\, ,
 |\vec{q}|=|\vec{q^*}|=\sqrt{{{q^*}^0}^2-M^2}.
\label{eq:momdimeson}
\ee

For the kaon loops with exchange of a $K^*$ vector meson,
 Fig.\ref{fig:VMD_loop_kaons_w},
 the corresponding expression for the
$t$-matrix,  is

\begin{equation}
t^{\mu\nu} =\frac{G \overline{G}}{2\sqrt{2}}
\,\bar{t}'^{\mu\nu}\, \frac{4}{\sqrt{3}} t_{KK,\pi\pi}^{I=0}.
\end{equation}
where we must bear in mind that $\bar{t}'^{\mu\nu}$ has the same
structure as in Eq.~(\ref{eq:tbar}) but the masses and widths of
the particles are correspondingly changed.

In an analogous way, we can evaluate the same kind of diagrams but
with an intermediate axial-vector meson  instead of a vector meson both for
pion and kaon intermediate loops. The diagrams are thus the same as
Fig.~\ref{fig:VMD_loop_pions_w} but substituting $\rho$ by $b_1$ and the
same as Fig.~\ref{fig:VMD_loop_kaons_w} substituting $K^*$ by $K_1(1270)$
and $K_1(1400)$.
Given the different momentum structure of Eq.~(\ref{eq:tAtree}) with
respect to Eq.~(\ref{eq:tVtree}), the $t$-matrix is slightly different,
giving

\begin{equation}
t^{\mu\nu} = \frac{4\sqrt{2} D\overline{D}}{M m_\omega m_b^2}
\left[(q^*\cdot q g^{\mu\nu}-q^{\mu} {q^*}^{\nu}) G_{\pi\pi}
-\tilde{t}'^{\mu\nu}\right] 2 t_{\pi\pi,\pi\pi}^{I=0}.
\label{eq:kkaa}
\end{equation}

Given the Lorentz covariance of the factor 
$(q^*\cdot q g^{\mu\nu}-q^{\mu}\cdot{q^*}^{\nu})$,
coming from the 
$(\epsilon^*\cdot\epsilon \ q^*\cdot q-\epsilon^*\cdot q \ \epsilon\cdot q^*)$
term in Eq.~(\ref{eq:tAtree}), it can be already evaluated with
the momentum variables of the 
$\J$ rest frame, that is, $q^*=(M,0,0,0)$,
 $q^0=\frac{M^2+m_\omega^2-M_I^2}{2M}$,
  ${|\vec{q}|}=\sqrt{{q^0}^2-m_\omega^2}$.
For the $\tilde{t}'^{\mu\nu}$ part, which comes from the part of
Eq.~(\ref{eq:tAtree}) that contains a propagator, we still resort
to evaluate it in the dimeson rest frame and boost it to the $\J$
rest frame. Hence, by analogy to Eq.~(\ref{eq:tbar}), we have

\begin{equation}
\tilde{t}'^{\mu \nu} = \left(\begin{array}{cccc} 
0 & 0 & 0 & 0 \\
\frac{I_a+I_b-I_0}{M^2} |\vec q|({{q^*}^0}^2-{\vec{q}\,^2})
({{q^*}^0}- {q^0})
 & \frac{I_a+I_b-I_0}{M^2} ({{q^*}^0}^2-{\vec{q}\,^2})
 ({{q^*}^0} {q^0}-{\vec{q}\,^2}) & 0 & 0 \\
0 & 0 & \tilde{t}'_{22} & 0 \\
0 & 0 & 0 & \tilde{t}'_{33} \\
\end{array}
\right)
\label{eq:tprime}
\end{equation}
with 
$\tilde{t}'_{22}=\tilde{t}'_{33}\equiv -{q^0} {{q^*}^0} I_0 + |\vec q|({q^0}
+{{q^*}^0})I_1 + ({{q^*}^0} {q^0} - 2 {\vec{q}\,^2})I_a - {\vec{q}\,^2} I_b$,
where in this case the momenta appearing in the expression are those of
the dimeson rest frame, Eq.~(\ref{eq:momdimeson}).
As mentioned before, the $I_i$ integrals have to be evaluated with the appropriate
masses and widths of the corresponding mesons in the loops.

For the  diagrams with kaon loops and $K_1(1270)$ intermediate
exchange, the expression for the $t$-matrix, obtained in an analogous way,
is
\begin{equation}
t^{\mu\nu} = \frac{4}{M m_\omega m_{K_1(1270)}^2}
c\overline{D}\frac{1}{\sqrt{2}}(cD-sF)
\left[(q^*\cdot q g^{\mu\nu}-q^{\mu}{q^*}^{\nu}) G_{KK}
-\tilde{t}'^{\mu\nu}\right] \frac{4}{\sqrt{3}} t_{KK,\pi\pi}^{I=0}.
\label{eq:bb}
\end{equation}
In  Eq.~(\ref{eq:bb}), $c\equiv cos(\alpha)$ and $s\equiv
cos(\alpha)$ are the cosinus and sinus of the mixing angle,
$\alpha$, between the isospin $1/2$ members of the axial-vector
octets to give the physical $K_1(1270)$ and $K_1(1400)$ states,
Eq.~(\ref{eq:mixK}). For the diagrams with $K_1(1400)$
intermediate state the amplitude is the same but changing
$m_{K_1(1270)}\to m_{K_1(1400)}$, $F\to -F$, $c\to s$ and $s\to
c$ and replacing the masses and widths of the $K_1(1270)$
by those of the $K_1(1400)$ in the evaluation of
$\tilde{t}'^{\mu\nu}$.

The expressions for the amplitudes of all the mechanisms corresponding to
the other channels, ($\J\to\omega K \bar K$, $\phi\pi^+\pi^-$,
$\phi K \bar K$), are given in the Appendix.

\section{Results}

In the model described so far, the only unknown parameters are
the overall strength, $\widetilde{g}$, of the mechanisms of
Fig.~\ref{fig:chiral_loops} containing the direct $\J PPV$
vertex, (which we will call in what follows "direct"
terms), and the OZI rule violation parameter,
$\lambda_\phi$, (see Eq.~(\ref{eq:tchiral_loops})).  The other
constants and couplings appearing in the model are theoretically
fixed or obtained from direct decays with the PDG values, and
their experimental uncertainties will be taken into account when
evaluating the theoretical error bands of our results.
Therefore, there is no freedom in the mechanisms different to
"direct" terms in the sense that its strength and shape is fixed,
up to some sign which will be discussed below. Taken this into
account, the philosophy is to fit the full model to invariant
mass distributions of the $\J\to\omega\pi^+\pi^-$ and
$\J\to\phi\pi^+\pi^-$ decay channels to obtain the two
free parameters.

The experimental  data for $\J\to\omega\pi^+\pi^-$ is taken from BES
\cite{WuBES} and DM2  \cite{Augustin} experiments. 
For $\J\to\phi\pi^+\pi^-$ the
data has been taken from DM2 \cite{Falvard} and MARK-III
\cite{Lockman} experiments.
In most experiments the mass distributions are given in arbitrary
units. However, it is possible to find the absolute normalization
from information given in the papers or by using the branching
ratios for each channels from the PDG, given the fact that the
experiments provide the data in the full range of invariant mass
allowed. This we have done in the present work and is a novelty
with respect to former works on the issue. The fact that all
mechanisms in our approach, except the "direct" mechanisms, have
a fixed strength forces us to carry a fit to absolute
data to make meaningful the extracted values of the parameters of
the "direct" mechanisms.

Apart from the freedom due to the
$\widetilde{g}$ and $\lambda_\phi$
parameters, we have the uncertainty in  the signs of the $\J VP$
and $\J AP$ couplings, $\overline{G}$ and  $\overline{D}$
respectively. This uncertainty reflects in our model in only 
the relative sign between $\overline{G}$ and $\overline{D}$.
This is the case in the decays we are studying in the present
work since the relative sign with $\widetilde{g}$ is absorbed in
the $\widetilde{g}$ itself, which is a free parameter.
Therefore, given also the uncertainty in the overall sign of the
full amplitude, we will consider $\overline{D}$ to be positive
and will explicitly discuss the cases with  $\overline{G}>0$ and
$\overline{G}<0$.

We will consider invariant dimeson
masses up to $\sim 1200\MeV$ since
this is approximately  the maximum range of applicability of the
chiral unitary approach techniques used in the evaluation of the
meson-meson final state interaction throughout this work. In the
$\J\to\omega\pi^+\pi^-$ channel the tail of the $f_2(1270)$
meson  influences the region of high invariant masses and then
is the only source of background which is not generated by our
theoretical model. Therefore we have phenomenologically included
this resonance by fitting an $f_2(1270)$ Breit-Wigner,
convoluted by the $\J\to\omega\pi^+\pi^-$ phase space,  to the
$f_2(1270)$ peak (not shown in the figures) of the BES and DM2
data, and then we have added it to $d\Gamma/dM_I$. This
incoherent sum is accurate enough since the $f_2(1270)$ meson is
a D-wave and does not interfere with the scalar $f_0(980)$ which
dominates the process at these energies.

We have obtained  the two following results for the fits to the
$\J\to\omega\pi^+\pi^-$ and $\J\to\phi\pi^+\pi^-$ data:

\ba
\nonumber
\textrm{for }\overline{G}>0 \,\,:\quad
\widetilde{g}=0.032\pm 0.001 \quad;\lambda_\phi=0.12\pm0.03\quad;
(\frac{\chi^2}{d.o.f.}\simeq 3.4) \\
\textrm{for }\overline{G}<0 \,\,:\quad
\widetilde{g}=0.015\pm 0.001 \quad;\lambda_\phi=0.20\pm0.03\quad;
(\frac{\chi^2}{d.o.f.}\simeq 3.1)
\label{eq:fits}
\ea

The theoretical uncertainties given in the results of the fits in
Eq.~(\ref{eq:fits}) are a rough but safe estimates of the statistical
errors of the fit. Actually, we have obtained a strong correlation between
the two parameters, (given by the off-diagonal terms of the covariance
matrices), and therefore the uncertainties given in Eq.~(\ref{eq:fits})
are a conservative error bound extracted from the confidence ellipse of
the fit. The results obtained in Eq.~(\ref{eq:fits}) show clearly
$\lambda_\phi\ne 0$, and reasonably smaller than $1$, which quantifies the
OZI rule violation discussed in the present work. It is worth comparing
the $\lambda_\phi$ values obtained in the present work with the one
obtained in Ref.~\cite{OllerJ}, $\lambda_\phi=0.17\pm0.06$, which
falls in the middle of our two solutions. The new mechanisms that
we have introduced have definitely a relevant role in the
process, but it is also rewarding that in spite of the
uncertainties in the sign of $\overline{G}$, the values of
$\lambda_\phi$ obtained are qualitatively similar, and also
similar to the value found in Ref.~\cite{OllerJ}.  

\begin{figure}
\centerline{\protect\hbox{
\psfig{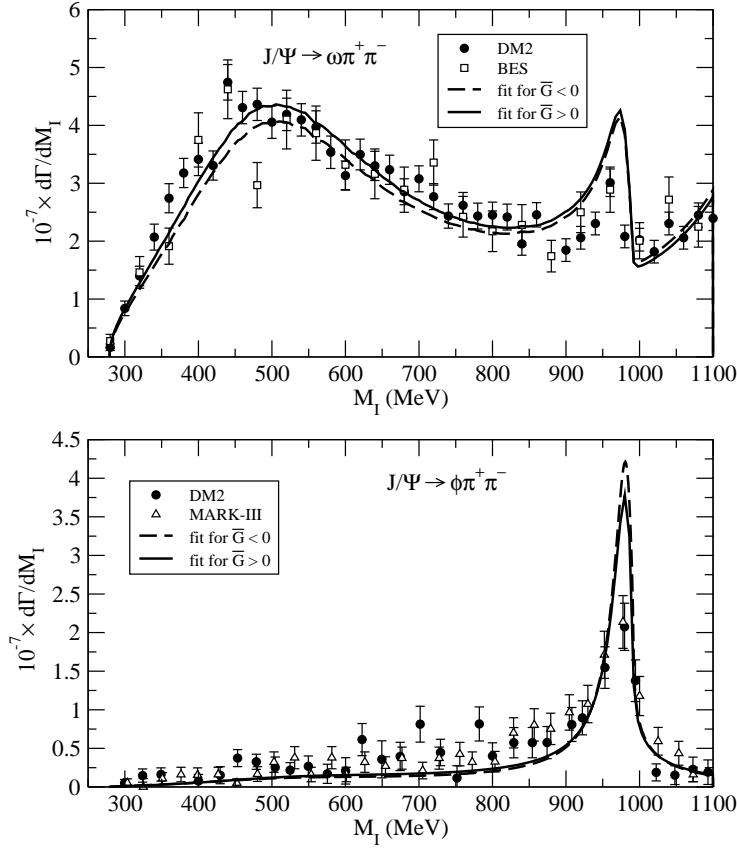}}}
\caption{Results of the fits to the invariant mass distribution of the
$\J\to\omega\pi^+\pi^-$ and $\J\to\phi\pi^+\pi^-$ decay channels. Solid
line: result for $\overline{G}>0$; dashed line: 
result for $\overline{G}<0$.}
\label{fig:res1}
\end{figure}

\begin{figure}
\centerline{\protect\hbox{
\psfig{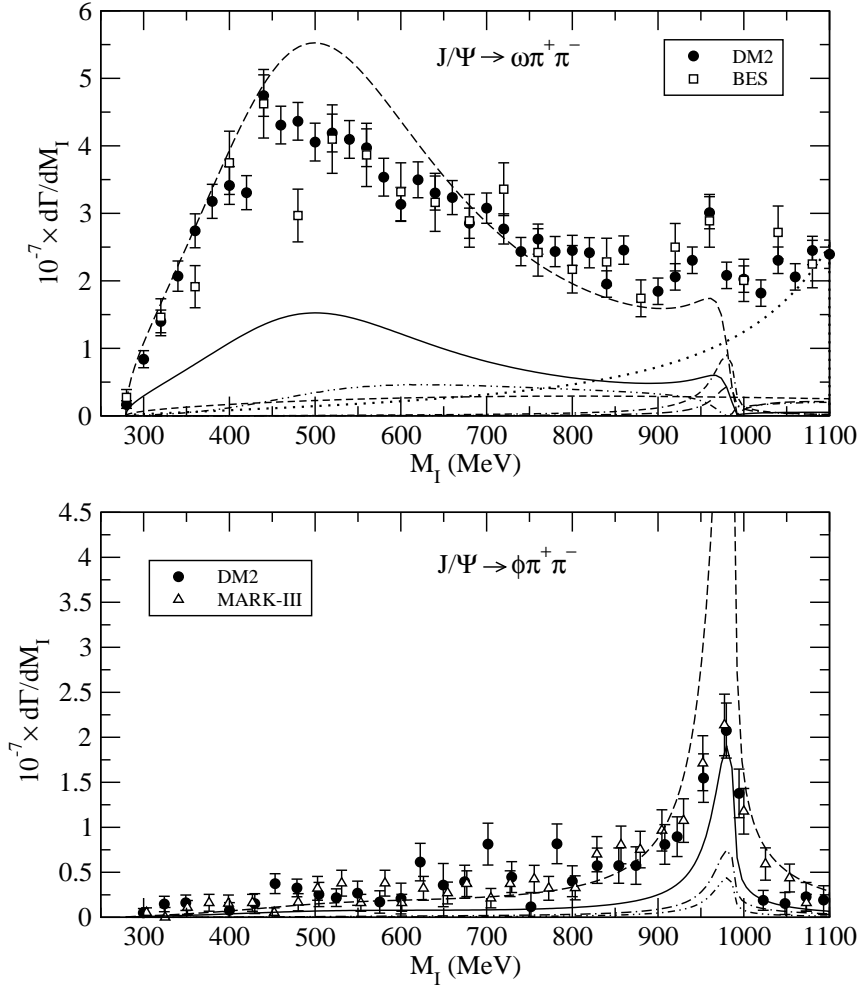}}}
\caption{Different contributions to the $\pi^+\pi^-$ invariant
mass distribution. The nomenclature of the different lines are
explained in the text.}
\label{fig:res2}
\end{figure}

In Fig.~\ref{fig:res1} we show the two results of the fits of
Eq.~(\ref{eq:fits}) for the invariant mass distribution of the
two pions in comparison with the experimental data. With the
solid line we show the solution of Eq.~(\ref{eq:fits}) for
$\overline{G}>0$ and with dashed line the solution for
$\overline{G}<0$. The theoretical curves have to be averaged
over the experimental bins since it can be specially important 
in the region of the $f_0(980)$ meson because of the narrowness
of the distribution. We have checked
that this bin average decreases the peak in the $f_0(980)$
region in the $\phi\pi\pi$ channel in around $15$\% and smooths
a little bit the curve at these masses in the $\omega\pi\pi$
channel, but we have not plotted it for simplicity. We observe
that there is a fair agreement with the experimental data for
both decay channels. Specially interesting is the good agreement
in the small bump  appearing  in the region of the $f_0(980)$
meson in the $\J\to\omega\pi^+\pi^-$  which had not been
considered before in previous theoretical works.  In
Fig.~\ref{fig:res1} we can see that both solutions give a very
similar final result, but the contributions of the various
mechanisms, specially the "direct" terms,
 is quite different in each case as it
can be seen in Fig.~\ref{fig:res2}. 
 In Fig.~\ref{fig:res2} we show the different contributions to
the invariant mass distribution coming from the various
mechanisms considered in our model. For the
$\J\to\omega\pi^+\pi^-$ channel the lines represent: "direct"
terms for $\overline{G}>0$  (long-dashed line), "direct" terms
 for
$\overline{G}<0$ (solid line), sequential vector meson exchange
at tree level (short-dashed line), pion loops of sequential
vector meson exchange (dash double-dotted line),  kaon loops of
sequential vector meson exchange (dash-dotted line), loops of
sequential exchange of $K_1(1270)$ (double-dash dotted line),
tail of the $f_1(1270)$ (dotted line). The intermediate $b_1$
meson contribution is too small to be visible in the plot and
then it has not been included in the figures. For the
$\J\to\phi\pi^+\pi^-$ channel the nomenclature of the lines is
the same but the double-dash dotted line represents the
$K_1(1400)$ exchange contribution,
the sequential vector meson exchange at tree level
and with pion loops are not plotted because they give a small
contribution and the $f_2(1270)$ does not give contribution. The
two solutions for the "direct" terms  (for $\overline{G}>0$ and
$\overline{G}<0$) look very different both in shape and in
strength, indicating the important role of the $\widetilde{g}$
and $\lambda_\phi$ parameters because of the strong and  non
trivial interferences of the "direct" terms with the other
mechanisms. In the $\omega\pi\pi$ channel, for $\overline{G}>0$
the interference with the rest of diagrams is mainly 
destructive and for $\overline{G}<0$ it is constructive. In the
$\phi\pi\pi$ channel the interferences are the other way
around. Specially crucial is the interference between all the
mechanisms in the $f_0(980)$ region, since many diagrams contribute
to it due to the final meson-meson state interaction. Therefore
it is not trivial to reproduce the good strength and shape in
the $f_0(980)$ region. It is important to stress again that
there is no freedom in the extra mechanisms besides the 
"direct" terms. Therefore, their strength and shape are crucial in
order to determine the free parameters of the "direct" terms when
the fit  with the full model to the mass distribution with
absolute normalization is performed.

Special attention and discussion deserves the low mass region in
the $\omega\pi\pi$ channel: the visible bump at  $\sim 500\MeV$
has been claimed in the literature to be a direct effect  of
the  $\sigma$ meson, but we will see that  one has to make a
very careful analysis if one wants to extract the physical
$\sigma$ mass and width from this reaction. In previous analyses
of the DM2 data \cite{Augustin,Ishida} the authors used
two s-wave
Breit-Wigner plus polynomial shapes ignoring their mutual
interference. In the analysis  of Ref.~\cite{WuBES} using the
BES data, a slightly modified BW shape was used but
without considering possible interferences with other terms. We
will see in the following that the correct shape of the bump at
lower energies comes mainly from a subtle interference of the
$t_{\pi\pi,\pi\pi}^{I=0}$ amplitudes of
Fig.~\ref{fig:chiral_loops}b) with the contact term of
Fig.~\ref{fig:chiral_loops}a).
\begin{figure}
\centerline{\protect\hbox{
\psfig{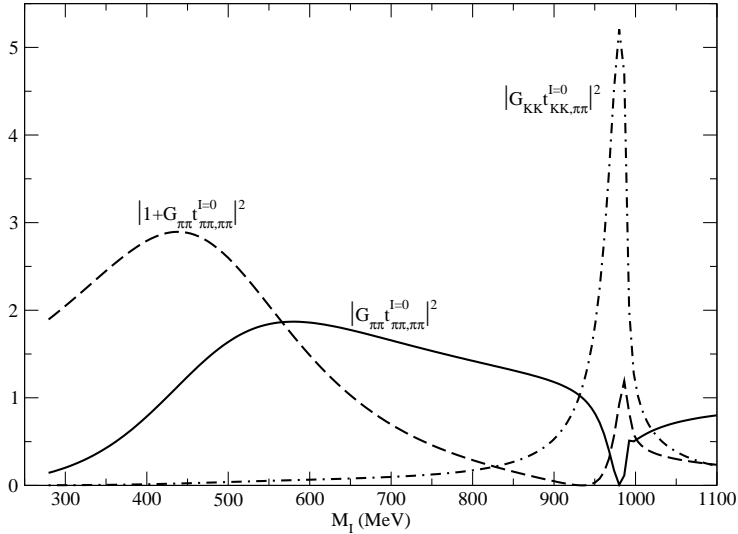}}}
\caption{Modulus square of the different pieces in
Eq.~(\ref{eq:tchiral_loops})
involving the two
meson loop and the meson-meson unitarized amplitude.}
\label{fig:1mast}
\end{figure}
In Fig.~\ref{fig:1mast} we plot the modulus squared of
$G_{KK}t_{KK,\pi\pi}^{I=0}$ (dashed-dotted line),
$G_{\pi\pi}t_{\pi\pi,\pi\pi}^{I=0}$ (solid line) and
$1+G_{\pi\pi}t_{\pi\pi,\pi\pi}^{I=0}$ (dashed line).
In the unitary chiral 
models, the scalar mesons appear as poles in the second Riemann sheet
of the $t_{MM,M'M'}$ scattering amplitude in the scalar-isoscalar channel 
\cite{Oller:1998hw,npa}.
Actually the $f_0(980)$ appears in the $987-i14\MeV$ position and the
$\sigma$ meson
in  $445-i221\MeV$, but its physical mass, given by the
distribution in the real axis, is around $600\MeV$ because of the large
width and the complicated distribution in the complex plane of the $\sigma$
meson pole
\cite{Oller:1998hw}.
Therefore the $\sigma$ bump would be seen
in around
$600\MeV$ if only the term containing the $t_{\pi\pi,\pi\pi}^{I=0}$
 amplitude would be present, this is, with a shape similar to the
 solid line in Fig.~\ref{fig:1mast}.
 When the $t_{\pi\pi,\pi\pi}^{I=0}$ term interferes
with the contact term of Fig.~\ref{fig:chiral_loops}a), represented by the
"$1$" in the formulae, the shape in the $\sigma$ region  is strongly modified  to
something much more similar to the final shape of the $d\Gamma/dM_I$
curve. Therefore one can conclude that when doing an analysis of the
experimental data to extract the $\sigma$ mass and width one has to allow
the $\sigma$ meson amplitude to interfere with a polynomial containing at
least  the constant term. A similar narrow $\pi^+\pi^-$
distribution is found experimentally in the $\J\to p\bar
p\pi^+\pi^-$ decay \cite{Eaton:1983kb} and the theoretical
explanation found in \cite{Li:2003zi} was analogous to the one
found here, from the interference of a tree level mechanism with
a rescattering mechanism involving the $\pi\pi$ scattering matrix
in the "$\sigma$" channel. This situation is different to the one
found in other reactions like $\gamma\gamma\to \pi^0\pi^0$
\cite{Oller:1997yg}, where the direct contact term is forbidden
and the amplitude is dominated by the loop terms proportional to
$t_{\pi\pi,\pi\pi}^{I=0}$. In this case the shape of the
$\pi^0\pi^0$ distribution is very wide, resembling the solid
line in Fig.~\ref{fig:1mast} 
\cite{Oller:1997yg,Marsiske:1990hx,Oest:1990ki}.
It is also interesting to present a different interpretation of
this peak. Since in the $\sigma$ region the $K\bar K$ channel is
not important we can use the Bethe-Salpeter equation with only
one channel, $\pi\pi$, and then we have
\be
t_{\pi\pi,\pi\pi}^{I=0}=V+VG_{\pi\pi}t_{\pi\pi,\pi\pi}^{I=0}
=V(1+G_{\pi\pi}t_{\pi\pi,\pi\pi}^{I=0}).
\ee
Since in the first of Eqs.~(\ref{eq:tchiral_loops}), neglecting
the $K \bar K$ channel, one has the contribution
$1+G_{\pi\pi}t_{\pi\pi,\pi\pi}^{I=0}$,
then one can make
\be
1+G_{\pi\pi}t_{\pi\pi,\pi\pi}^{I=0}
=\frac{t_{\pi\pi,\pi\pi}^{I=0}}{V}.
\ee
Hence, the shape in the $\J\to\omega\pi^+\pi^-$ channel around
$M_I=500\MeV$ is given roughly by 
$|t_{\pi\pi,\pi\pi}^{I=0}/V|^2$ and
it happens that $V\,(V=-(M_I^2-m_\pi^2/2)/f^2)$ is
more strongly
dependent on $M_I$ than $t_{\pi\pi,\pi\pi}^{I=0}$, it grows
faster as a
function of $M_I$ and the ratio  $t_{\pi\pi,\pi\pi}^{I=0}/V$ 
decreases faster
as a function of $M_I$ than $t_{\pi\pi,\pi\pi}^{I=0}$, producing this
apparent narrower peak, which does not reflect the $M_I$
 dependence
of the $t_{\pi\pi,\pi\pi}^{I=0}$ matrix but rather the $M_I$
 dependence
of the kernel $V$. 
In order to see the $t_{\pi\pi,\pi\pi}^{I=0}$
amplitude one has to resort to reactions where the Born term is
forbidden like in the $\gamma\gamma\to\pi^0\pi^0$ reaction.

On the other hand, by looking at Fig.~\ref{fig:1mast} and
Eq.~(\ref{eq:tchiral_loops}), one can understand  the different
weights of the $f_0(980)$ and $\sigma$ mesons in the different
decay channels. 
From Eq.~(\ref{eq:tchiral_loops}) one can see for
$\J\to\phi\pi^+\pi^-$ that the term generating the $\sigma$, 
$t_{\pi\pi,\pi\pi}^{I=0}$, has a coefficient $(\nu -1)$
which vanishes if the 
OZI rule is exact, as we pointed out before.
Given the smallness of the OZI rule breaking that we have found,
this term is small and hence there is practically no trace of the
$\sigma$ in the $\J\to\phi\pi^+\pi^-$ decay. On the other hand,
there are no OZI restrictions in the coefficient of the 
$t_{KK,\pi\pi}^{I=0}$ amplitude, which contains the $f_0(980)$
pole, and hence the $f_0(980)$ resonance appears neatly and
dominates the distribution.
 In the
$\omega\pi^+\pi^-$ channel, the $t_{KK,\pi\pi}^{I=0}$
and  $t_{\pi\pi,\pi\pi}^{I=0}$ terms are both not OZI suppressed
and have comparable weight and hence both the $\sigma$ and
the $f_0(980)$ show up with comparable strength.
 This discussion is only approximate,
in the sense that the final shapes and strengths are determined
when added coherently to the rest of mechanisms of the model,
but it describes accurately the qualitative behavior.

\begin{figure}
\centerline{\protect\hbox{
\psfig{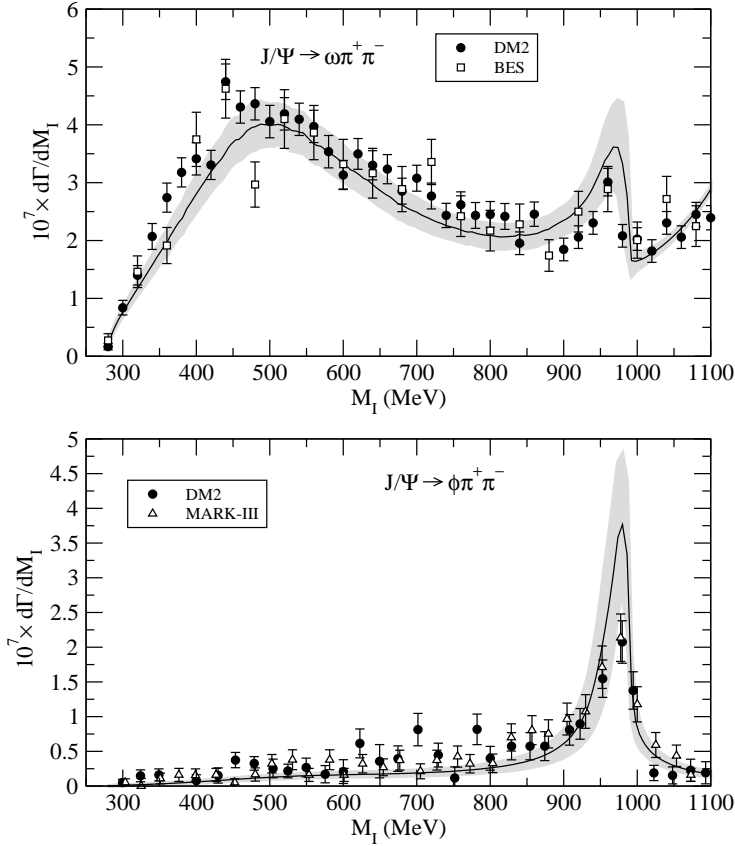}}}
\caption{Final result with the theoretical error bands.}
\label{fig:res5}
\end{figure}

In order to give an idea of the uncertainties of our model, we show in
Fig.~\ref{fig:res5} the final results with the theoretical error band 
obtaining implementing a Montecarlo gaussian sampling of the parameters
used in the model
within their error bounds.
 We only show the result for $\overline G<0$,
since the result for $\overline G>0$ produces a very similar plot.
The agreement within errors is quite fair even in the $f_0(980)$ in the
$\J\to\phi\pi^+\pi^-$ channel if one would reduce by an extra
$\sim 15$\% the plot in
the $f_0(980)$ region due to the experimental binning.

With the results obtained so far, it is interesting to test the model in
other decay channels having $K^+K^-$ as final pseudoscalar pair,
without introducing any extra freedom.
We have thus evaluated the invariant mas distribution of the
$K^+K^-$
in $\J\to\omega K^+K^-$ and $\J\to\phi K^+K^-$ decay channels.
\begin{figure}
\centerline{\protect\hbox{
\psfig{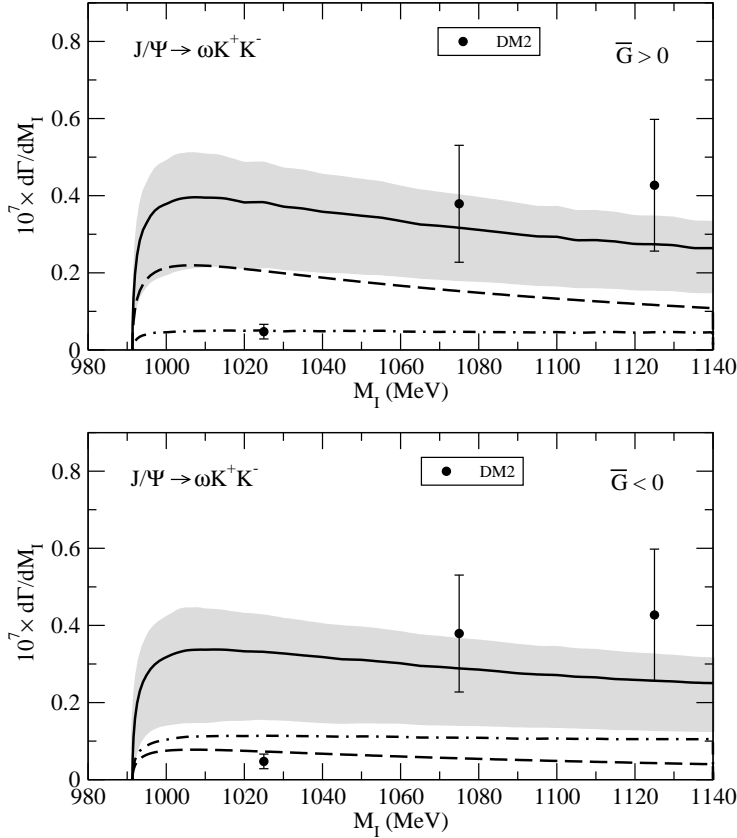}}}
\caption{Results for the $K^+K^-$ invariant mass distribution
 for the $\J\to\omega K^+K^-$ for the two different results of
 Eq.~(\ref{eq:fits}). Solid line: full model with the theoretical error
 band; dashed line: "direct" terms, dashed-dotted line: the rest of
 mechanisms.}
\label{fig:res6}
\end{figure}
\begin{figure}
\centerline{\protect\hbox{
\psfig{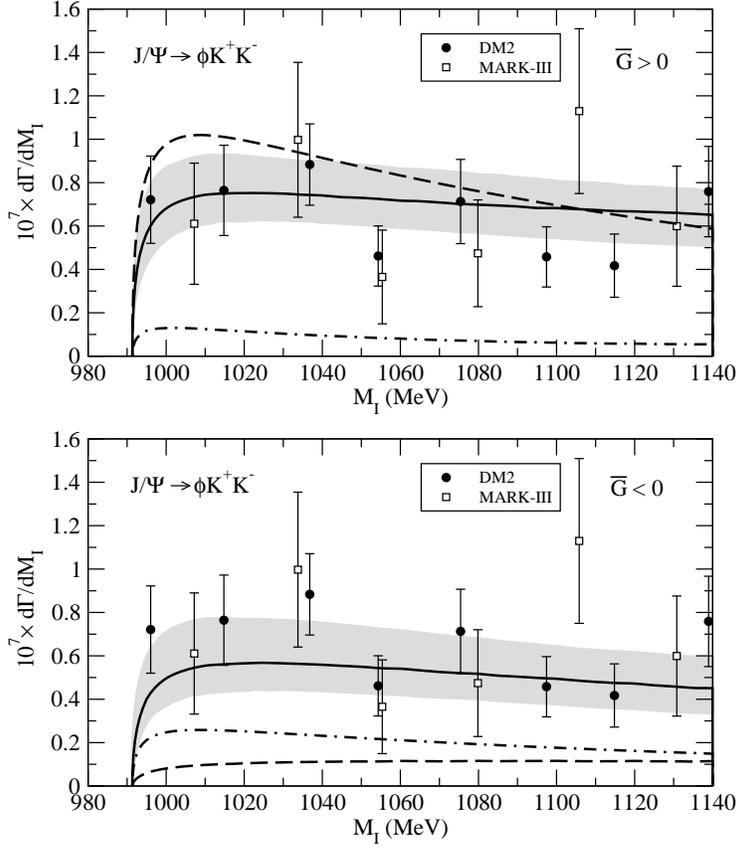}}}
\caption{Results for the $K^+K^-$ invariant mass distribution
 for the $\J\to\phi K^+K^-$ for the two different results of
 Eq.~(\ref{eq:fits}). Solid line: full model with the theoretical error
 band; dashed line: "direct" terms, dashed-dotted line: the rest of
 mechanisms.}
\label{fig:res7}
\end{figure}
The experimental data for the $\J\to\omega K^+K^-$ decay channel 
has been taken from
DM2 \cite{Falvard} experiment.
The experimental data $\J\to\phi K^+K^-$ has been taken from
DM2 \cite{Falvard} and MARK-III \cite{Lockman} experiments.
There are large  uncertainties in the total normalization
of the experimental data given the smallness of the phase space region
considered. Actually one has to consider an extra $\sim 30$\% of
systematic error in the experimental data for the $\omega K^+K^-$ channel
\cite{Falvard} not included in the data shown.
We present in Figs.~\ref{fig:res6} and \ref{fig:res7}
the results for both solutions of
Eq.~(\ref{eq:fits}) showing the contribution of the "direct"
mechanisms (dashed line),
all the rest of diagrams together (dashed-dotted line)
and the final result with the theoretical error band (solid line).
It is important to stress that nothing has been fitted in these
 channels,
meaning that we are using the same results of Eq.~(\ref{eq:fits}).
 We see again the important role played by the interferences
between the "direct" terms and the rest of diagrams in order to obtain the
final shape and strength of the invariant mass distributions.
In the plots of both channels one can see the
 trend of the curve accumulating strength close to the threshold,
 given the proximity of the $f_0(980)$ resonance below threshold.
 The agreement of the results is quite fair, giving the large experimental
uncertainties, also in the normalization,
and the non trivial $f_0(980)$ meson effect. 

It is remarkable that the strength of the ·direct" terms is
very different in these cases. It is very large for the case
$\overline{G}>0$ and quite small for $\overline{G}<0$,
particularly in the $\J\to \phi K^+K^-$ channel, in spite of 
which the final results are very similar and this experiment
does not help to discriminate between both solutions. The study
of these $K^+K^-$ decay processes shows however how important
the "non direct" mechanisms are for these processes.


\section{Conclusions}

We have made a comprehensive study of the $\J$ decay into one
vector meson and two pseudoscalars, addressing mainly questions
about the scalar mesons and the role played by the OZI rule. 
Apart from "direct" mechanisms used in previous works, we have
included other mechanisms which proved relevant in similar
reactions. The first and most important mechanism considered is
the one containing the "direct" $\J VPP$ vertex implementing
also the final state interaction of the pseudoscalar pair,
following the techniques of the chiral unitary approach, which
allows to extend the predictive power of ChPT up to energies 
$\sim 1.2\textrm{ GeV}$.
This chiral unitary approach implements unitarity in coupled
channels and many resonances, specially the scalar mesons,  are
generated dynamically, showing up as poles in the meson-meson
scattering amplitudes.  The $\J VPP$ vertex amplitudes have been
constructed using $SU(3)$ arguments to relate the different
channels and also parametrizing the amplitudes in a way which 
clearly manifest the role of the OZI rule. We have
also included in the model other mechanisms where the $\J$
decays into a vector or axial-vector and a
pseudoscalar meson and vector or
axial-vector meson subsequently decays
into the final vector and another
pseudoscalar meson. For the $\J$ to vector and axial-vector
coupling, we  have proposed a suitable phenomenological
Lagrangian. We have also implemented  in this mechanisms the
final meson-meson state interaction which has turned out to be
important in our results, since it dynamically generates
 the scalar mesons. These mechanisms are
crucial to determine the final shape and strength of the
invariant mass distribution through interferences with the
"direct" terms. Given the fact that the strength of these
sequential vector and axial-vector exchange mechanisms is fixed,
it was important to carry a fit to the data with absolute
normalization in order to obtain meaningful values for the
parameters of the "direct" terms.

The only two free parameters in our model are the coupling of
the direct $\J VPP$ vertex and the OZI rule violation parameter,
$\lambda_\phi$. Fitting our model to the $\J\to\omega\pi^+\pi^-$
and $\J\to\phi\pi^+\pi^-$ experimental data we obtain values of
$\lambda_\phi$ clearly different from zero and reasonably
smaller than one, what manifests the OZI rule violation within
reasonable values.

 Concerning the scalar mesons, it is important to stress first
that the final shape and strength of the bump appearing in the
$\pi\pi$ invariant mass distribution in the $\omega\pi\pi$
channel at $\sim 500\MeV$ is determined by a subtle interference
between the final state interaction  in the direct $\J VPP$
mechanisms and the tree level direct $\J VPP$ decay. This means
that the shape and position of the bump does not directly
represent the physical properties of the $\sigma$ meson, since
it is distorted due to interferences with other terms not
related to the $\sigma$ meson.
Alternatively, we also showed that the relative narrowness of the
peak was due to the stronger $M_I$ dependence of the kernel $V$ of
the Bethe-Salpeter equation rather than the $M_I$ dependence of the
$t_{\pi\pi,\pi\pi}^{I=0}$ scattering amplitude.
 Therefore one has to be
extremely careful when using the experimental
 data to extract the physical
$\sigma$ meson properties by fitting Breit-Wigner-like shapes.

On the other hand, the relative weights of the $f_0(980)$  and
the $\sigma$ meson are well reproduced in both the 
$\omega\pi\pi$ and $\phi\pi\pi$ channels in spite of their large
difference in these channels. This relative weight is mainly
determined by the OZI rule violation parameter and the
interferences of the "direct" terms with the other mechanisms,
specially in the $f_0(980)$ region. In our model, since the
scalar mesons are dynamically generated through the resummed
meson-meson amplitude, the relative weight between the
$f_0(980)$ and the $\sigma$ mesons is related to the relative
weight between the $K\bar K\to\pi\pi$ and $\pi\pi\to\pi\pi$, in
$I=0$, scattering amplitudes. Specially remarkable is the
agreement in the $f_0(980)$ region of the $\omega\pi\pi$ channel
despite the smallness of the bump.

Finally, we have applied our results to the  $\omega K\bar K$
and $\phi K\bar K$ decay channels, obtaining a fair agreement
without introducing any extra freedom in the model. This is a
nice test of the present model, both reproducing the absolute
strength, and also the shape, which shows much strength close to
threshold as a reflection of the proximity of the $f_0(980)$
resonance below threshold.

In conclusion, we have obtained a good description of these
interesting $\J$ decays combining phenomenological Lagrangians
and the techniques of the chiral unitary approach to implement
the final state rescattering of the pseudoscalar pairs,
quantifying the controversial non-trivial role of the scalar
mesons and the violation of the OZI rule.
The fact that once more one is able to reproduce the shape and
strength of the $f_0(980)$ and the $\sigma$ resonances without the
need to introduce them as explicit degrees of freedom provides an
extra support to the idea of the nature of these resonances as
dynamically generated from the interaction of the mesons.

\section*{Acknowledgments}

Two of us, J.E.P. and L.R., acknowledge support from the
Ministerio de Educaci\'on, Cultura y Deportes. 
This work is
partly supported by DGICYT contract number BFM2003-00856,
and the E.U. EURIDICE network contract no. HPRN-CT-2002-00311.

\vspace{2cm}

\begin{center}
{\bf \Large{Appendix:
Amplitudes for all the decay channels}}\\
\end{center}

{\it \large{A.1: Meson loops from direct $J/\Psi PPV$ vertex}} \\

The amplitudes for these mechanisms for all the channels are given in 
Eq.~(\ref{eq:tchiral_loops})  \\

{\it \large{A.2: Sequential vector meson exchange: tree level}} \\

- $J/\Psi\to\omega \pi^+\pi^-$:

\begin{equation}
t=-\frac{G\overline{G}}{\sqrt{2}}
\left[\frac{P^2\{a\}+\{b(P)\}}{M_{\rho}^2-P^2-iM_\rho
\Gamma_\rho(P^2)}
+\frac{P'^2\{a\}+\{b(P')\}}{M_{\rho}^2-P'^2-iM_\rho
\Gamma_\rho(P'^2)}
\right]
\end{equation}
where 
 $P=p_1+q$, $P'=p_2+q$ and
\begin{eqnarray} \nonumber
\{a\}&=&\epsilon^*\cdot\epsilon \ q^*\cdot q
- \epsilon^*\cdot q \ \epsilon\cdot q^* \\
\{b(P)\}&=&-\epsilon^*\cdot\epsilon \ q^*\cdot P \ q\cdot P
- \epsilon\cdot P \ \epsilon^*\cdot P \ q^*\cdot q
+ \epsilon^*\cdot q \ \epsilon\cdot P \ q^*\cdot P
+  \epsilon\cdot q^* \ \epsilon^*\cdot P \ q\cdot P
\nonumber
\end{eqnarray}

- $J/\Psi\to\phi \pi^+\pi^-$: 

\be
 t=0
\ee

- $J/\Psi\to\omega K^+K^-$: \\

The diagrams are like in Fig.~\ref{fig:VMD_tree_w} channel 
 but changing $\rho$ by
$K^*$ and pions by kaons.

\begin{equation}
t=-\frac{G\overline{G}}{2\sqrt{2}}
\left[\frac{P^2\{a\}+\{b(P)\}}{M_{K^*}^2-P^2-iM_{K^*}
\Gamma_{K^*}(P^2)}
+\frac{P'^2\{a\}+\{b(P')\}}{M_{K^*}^2-P'^2-iM_{K^*}
\Gamma_{K^*}(P'^2)}
\right]
\end{equation}
\\

- $J/\Psi\to\phi K^+K^-$:

The diagrams are like in the $\omega K^+K^-$ case changing $\omega$ by $\phi$.

\begin{equation}
t=-\frac{G\overline{G}}{2}
\left[\frac{P^2\{a\}+\{b(P)\}}{M_{K^*}^2-P^2-iM_{K^*}
\Gamma_{K^*}(P^2)}
+\frac{P'^2\{a\}+\{b(P')\}}{M_{K^*}^2-P'^2-iM_{K^*}
\Gamma_{K^*}(P'^2)}
\right]
\end{equation}
\\

{\it \large{A.3: Sequential axial-vector meson exchange: tree level}} \\

- $J/\Psi\to\omega \pi^+\pi^-$:

\begin{eqnarray}
t &=& \frac{4\sqrt{2}D\overline{D}}{M m_\omega m_b^2} 
\left[ (\epsilon^*\cdot\epsilon \ q^*\cdot q
- \epsilon^*\cdot q \ \epsilon\cdot q^*) \ + \ 
 \frac{1}{M_{b}^2-P^2-iM_b\Gamma_b(P^2)}  \cdot \right.\\ 
&\cdot& \left. \left(\epsilon^*\cdot\epsilon \ q^*\cdot P \ q\cdot P
+ \epsilon\cdot P \ \epsilon^*\cdot P \ q^*\cdot q
- \epsilon^*\cdot q \ \epsilon\cdot P \ q^*\cdot P
-  \epsilon\cdot q^* \ \epsilon^*\cdot P \ q\cdot P
\right)\right]
\nonumber
\end{eqnarray}

plus the crossed one, with $P'$.
\\

- $J/\Psi\to\phi \pi^+\pi^-$: 

\be
t=0
\ee

- $J/\Psi\to \omega K^+K^-$:

Intermediate $K_1(1270)$:

The diagrams are like in Fig.~\ref{fig:VMD_tree_w} changing $\rho$ by
$K_1(1270)$ and pions by kaons.

\begin{eqnarray}
\nonumber
t &=& \frac{4c\overline{D}(cD-sF)}{\sqrt{2}M m_\omega
 m_{K_1(1270)}^2} 
\left[ (\epsilon^*\cdot\epsilon \ q^*\cdot q
- \epsilon^*\cdot q \ \epsilon\cdot q^*) \ + \ 
 \frac{1}{M_{K_1(1270)}^2-P^2-iM_{K_1(1270)}
 \Gamma_{K_1(1270)}(P^2)}  \cdot \right.\\ 
&\cdot& \left. \left(\epsilon^*\cdot\epsilon \ q^*\cdot P \ q\cdot P
+ \epsilon\cdot P \ \epsilon^*\cdot P \ q^*\cdot q
- \epsilon^*\cdot q \ \epsilon\cdot P \ q^*\cdot P
-  \epsilon\cdot q^* \ \epsilon^*\cdot P \ q\cdot P
\right)\right]
\end{eqnarray}

plus the crossed one, with $P'$.
$c\equiv \cos{\alpha}$, $s\equiv \sin{\alpha}$. \\

Intermediate $K_1(1400)$:

  All the diagrams with intermediate $K_1(1270)$ also have the
  corresponding one with $K_1(1400)$. The amplitudes are the same but
changing $m_{K_1(1270)}\to m_{K_1(1400)}$,
$F\to -F$, $c\to s$ and $s\to c$. Therefore we will not explicitly give in
what follows the $K_1(1400)$ case. \\

- $J/\Psi\to \phi K^+K^-$:

\begin{eqnarray}
\nonumber
t &=& \frac{4c\overline{D}(cD-sF)}{M m_\phi
 m_{K_1(1270)}^2} 
\left[ (\epsilon^*\cdot\epsilon \ q^*\cdot q
- \epsilon^*\cdot q \ \epsilon\cdot q^*) \ + \ 
 \frac{1}{M_{K_1(1270)}^2-P^2-iM_{K_1(1270)}
 \Gamma_{K_1(1270)}(P^2)}  \cdot \right.\\ 
&\cdot& \left. \left(\epsilon^*\cdot\epsilon \ q^*\cdot P \ q\cdot P
+ \epsilon\cdot P \ \epsilon^*\cdot P \ q^*\cdot q
- \epsilon^*\cdot q \ \epsilon\cdot P \ q^*\cdot P
-  \epsilon\cdot q^* \ \epsilon^*\cdot P \ q\cdot P
\right)\right]
\end{eqnarray}
\\

{\it{\large A.4: Loops of sequential vector meson exchange}} \\

- $J/\Psi\to \omega \pi^+\pi^-$ \\

{\it Pion loops}:

\begin{equation}
t^{\mu\nu} = \frac{G \overline{G}}{\sqrt{2}}
\, \bar{t}'^{\mu\nu} \, 2 t_{\pi\pi,\pi\pi}^{I=0}
\end{equation}

$\bar{t}'^{\mu\nu}$ is given in Eq.~(\ref{eq:tbar}), but using
proper masses and widths in the evaluation. \\

{\it Kaon loops}:

\begin{equation}
t^{\mu\nu} =\frac{G \overline{G}}{2\sqrt{2}}
\, \bar{t}'^{\mu\nu} \, \frac{4}{\sqrt{3}} t_{KK,\pi\pi}^{I=0}
\end{equation}
\\

- $J/\Psi\to \phi\pi^+\pi^-$: \\

{\it Pion loops}:

\be
t^{\mu\nu} = 0
\ee

{\it Kaon loops}:

The diagrams are like in Fig.~\ref{fig:VMD_loop_kaons_w} changing $\omega$ by
$\phi$.

\begin{equation}
t^{\mu\nu} =\frac{G \overline{G}}{2}
\, \bar{t}'^{\mu\nu} \, \frac{4}{\sqrt{3}} t_{KK,\pi\pi}^{I=0}
\end{equation}

- $J/\Psi\to \omega K^+K^-$: \\

{\it Pion loops}:

The diagrams are like in Fig.~\ref{fig:VMD_loop_pions_w} changing the
final $\pi^+\pi^-$ by $K^+ K^-$.

\begin{equation}
t^{\mu\nu} = \frac{G \overline{G}}{\sqrt{2}}
\, \bar{t}'^{\mu\nu} \,
\sqrt{\frac{3}{2}} t_{\pi\pi,KK}^{I=0}
\end{equation}

{\it Kaon loops}:

\begin{equation}
t^{\mu\nu} =\frac{G \overline{G}}{2\sqrt{2}}
\, \bar{t}'^{\mu\nu} \,
2 t_{KK,KK}^{I=0}
\end{equation}

- $J/\Psi\to \phi K^+K^-$: \\

{\it Pion loops}:

\begin{equation}
t^{\mu\nu} = 0
\end{equation}

{\it Kaon loops}:

The diagrams are like in the $\omega K^+K^-$ case changing $\omega$ by
$\phi$.

\begin{equation}
t^{\mu\nu} =\frac{G \overline{G}}{2}
\, \bar{t}'^{\mu\nu} \,
2 t_{KK,KK}^{I=0}
\end{equation}
\\

{\it {\large A.5: Loops of sequential axial-vector meson exchange}} \\

- $J/\Psi\to \omega\pi^+\pi^-$: \\

{\it Pion loops}:

\begin{equation}
t^{\mu\nu} = \frac{4\sqrt{2} D\overline{D}}{M m_\omega m_b^2}
\left[(q^*\cdot q g^{\mu\nu}-q^{\mu} {q^*}^{\nu}) G_{\pi\pi}
-\tilde{t}'^{\mu\nu}\right] 2 t_{\pi\pi,\pi\pi}^{I=0}
\label{eq:aa}
\end{equation}

See subsection~\ref{sec:VAloops} for definition  of $\tilde{t}'^{\mu\nu}$ and
the momenta.
\\

{\it Kaon loops}:

\begin{equation}
t^{\mu\nu} = \frac{4}{M m_\omega m_{K_1(1270)}^2}
c\overline{D}\frac{1}{\sqrt{2}}(cD-sF)
\left[(q^*\cdot q g^{\mu\nu}-q^{\mu}{q^*}^{\nu}) G_{KK}
-\tilde{t}'^{\mu\nu}\right] \frac{4}{\sqrt{3}} t_{KK,\pi\pi}^{I=0}
\label{eq:bbb}
\end{equation}

- $J/\Psi\to\phi \pi^+\pi^-$ \\

{\it Pion loops}:

\begin{equation}
t^{\mu\nu} = 0
\end{equation}

{\it Kaon loops}: \\

Intermediate $K_1(1270)$ meson:

The diagrams are like in Fig.~\ref{fig:VMD_loop_kaons_w} changing $K^*$ by
$K_1(1270)$ and $\omega$ by $\phi$.

\begin{equation}
t^{\mu\nu} = \frac{4}{M m_\phi m_{K_1(1270)}^2}
c\overline{D}(cD+sF)
\left[(q^*\cdot q g^{\mu\nu}-q^{\mu}\cdot {q^*}^{\nu}) G_{KK}
-\tilde{t}'^{\mu\nu}\right] \frac{4}{\sqrt{3}} t_{KK,\pi\pi}^{I=0}
\label{eq:cc}
\end{equation}

- $J/\Psi\to\omega K^+K^-$: \\

{\it Pion loops}:

The diagrams are like in Fig.~\ref{fig:VMD_loop_pions_w} changing 
$\rho$ by $b_1$ and the
final $\pi^+\pi^-$ by $K^+ K^-$.

\begin{equation}
t^{\mu\nu} = \frac{4\sqrt{2} D\overline{D}}{M m_\omega m_b^2}
\left[(q^*\cdot q g^{\mu\nu}-q^{\mu} {q^*}^{\nu}) G_{\pi\pi}
-\tilde{t}'^{\mu\nu}\right] \sqrt{3}t_{\pi\pi,KK}^{I=0}
\label{eq:aaaa}
\end{equation}

{\it Kaon loops}:

The diagrams are like in Fig.~\ref{fig:VMD_loop_kaons_w} changing 
$K^*$ by $K_1(1270)$ and the
final $\pi^+\pi^-$ by $K^+ K^-$.

\begin{equation}
t^{\mu\nu} = \frac{4}{M m_\omega m_{K_1(1270)}^2}
c\overline{D}\frac{1}{\sqrt{2}}(cD-sF)
\left[(q^*\cdot q g^{\mu\nu}-q^{\mu}{q^*}^{\nu}) G_{KK}
-\tilde{t}'^{\mu\nu}\right] 2t_{KK,KK}^{I=0}
\label{eq:bbbb}
\end{equation}

- $J/\Psi\to \phi K^+K^-$: \\

{\it Pion loops}:

\be
t^{\mu\nu}=0
\ee

{\it Kaon loops}:

The diagrams are like in the $\omega K^+K^-$ case changing $\omega$ by
$\phi$.

\begin{equation}
t^{\mu\nu} = \frac{4}{M m_\phi m_{K_1(1270)}^2}
c\overline{D}(cD+sF)
\left[(q^*\cdot q g^{\mu\nu}-q^{\mu}\cdot {q^*}^{\nu}) G_{KK}
-\tilde{t}'^{\mu\nu}\right] 2t_{KK,KK}^{I=0}
\label{eq:cccc}
\end{equation}


\end{document}